\documentclass[preprint2]{aastex631}
\usepackage{xfrac}
\usepackage{amsmath,amssymb}
\usepackage{upgreek}
\usepackage{isomath}
\usepackage{rotating}

\newcommand\Rm{{\rm Rm} }
\newcommand\Rey{{\rm Re} }
\newcommand\Pm{{\rm Pm} }
\newcommand\kf{k_{\rm f} }
\newcommand\lf{l_{\rm f}}
\newcommand\SNr{\dot\sigma_{\rm Sn}}

\newcommand\EST{E_{\rm th}}

\newcommand{\vect}[1]{{{\mbox{\boldmath $#1$}}}}%also makes bold Greek letters
\newcommand{\mathbfss}[1]{\textbf{\textsf{#1}}}

\newcommand\cmcube{~ {\rm cm^{-3}}}
\newcommand\cs{ c_{\rm s}}
\newcommand\cplocal{ c_{\rm p}}

\newcommand\cv{ c_{ v}}

\newcommand\kpc{~ {\rm kpc}}
\newcommand\pc{~ {\rm pc}}
\newcommand\dx{ {\delta x}}

\newcommand\Myr{~ {\rm Myr}}

\newcommand\erg{~ {\rm erg}}
\newcommand\kms{~ {\rm km~s}^{-1}}

\newcommand\eKt{{\overline{e_{\rm K}}}}

\newcommand\NAA{{N128}}
\newcommand\NBA{{N256}}
\newcommand\NCA{{N512}}
\newcommand\PCB{{P512}}
\newcommand\PDC{{P1024}}
\newcommand\PEE{{P2048}}

\definecolor{midblue}{rgb}{0.0,0.4,0.7}
\definecolor{midgreen}{rgb}{0.1,0.6,0.3}
\definecolor{mypurple}{rgb}{0.7,0.3,0.8}

\newcommand{\rev}[1]{\textcolor{black}{#1}}

\received{\today}
\revised{\today}
\accepted{}
\submitjournal{ApJL}

\shorttitle{Magnetic Prandtl number sensitivity of small-scale dynamo}
\shortauthors{Gent et al.}
\begin{document}

\title{Asymptotic behaviour of galactic small-scale dynamos at modest magnetic Prandtl number}

\correspondingauthor{Frederick A. Gent}
\email{Email: frederick.gent@su.se, mordecai@amnh.org,\\
maarit.korpi-lagg@aalto.fi, touko.puro@aalto.fi,\\ matthias.rheinhardt@aalto.fi}

\author[0000-0002-1331-2260]{Frederick A. Gent}
\affiliation{Nordita, KTH Royal Institute of Technology and Stockholm University, Hannes Alfv\'ens v\"ag 12, Stockholm,
      SE-106 91 Stockholm, Sweden
}
\affiliation{
HPCLab, Department of Computer Science, Aalto University, P.O. Box 15400, FI-00076 Espoo, Finland
 }
\affiliation{
    School of Mathematics, Statistics and Physics,
      Newcastle University, NE1 7RU, UK
 }
\author[0000-0003-0064-4060]{Mordecai-Mark {Mac Low}}
\affiliation{
  Department of Astrophysics, American Museum of Natural History,
  %{200 Central Park West,}
  New York, NY 10024, USA
}
\author[0000-0002-9614-2200]{Maarit J. {Korpi-Lagg}}
\affiliation{
HPCLab, Department of Computer Science, Aalto University, P.O. Box 15400, FI-00076 Espoo, Finland
}
\affiliation{Nordita, KTH Royal Institute of Technology and Stockholm University, Hannes Alfv\'ens v\"ag 12, Stockholm,
      SE-106 91 Stockholm, Sweden
}
\author[0009-0008-8632-0385]{Touko {Puro}}
\affiliation{
HPCLab, Department of Computer Science, Aalto University, P.O. Box 15400, FI-00076 Espoo, Finland
}
\author[0000-0001-9840-5986]{Matthias {Rheinhardt}}
\affiliation{
HPCLab, Department of Computer Science, Aalto University, P.O. Box 15400, FI-00076 Espoo, Finland
}


\begin{abstract}
Magnetic fields are critical at many scales to galactic dynamics and structure,
including multiphase pressure balance, dust processing, and star formation.
Dynamo action determines their dynamical structure and strength.  Simulations
of combined large- and small-scale dynamos have successfully developed mean
fields with strength and topology consistent with observations but with
turbulent fields much weaker than observed, while simulations of small-scale
dynamos with parameters relevant to the interstellar medium yield turbulent
fields an order of magnitude below the values observed or expected
theoretically.  We use the {\sc Pencil Code} accelerated on GPUs with {\sc
Astaroth} to perform high-resolution simulations of a supernova-driven galactic
dynamo including heating and cooling in a periodic domain. Our models show that
the strength of the turbulent field produced by the small-scale dynamo
approaches an asymptote at only modest magnetic Prandtl numbers.  This allows
us to use these models to suggest the essential characteristics of this
constituent of the magnetic field for inclusion in global galactic models.  The
asymptotic limit occurs already at magnetic Prandtl number of only a few
hundred, many orders of magnitude below physical values in  the interstellar
medium and consistent with previous findings for isothermal compressible flows.
\end{abstract}

\section{Introduction}\label{sec:intro}
%=============================================================================

We report on the results of unprecedentedly high-resolution simulations of
supernova (SN)-driven turbulence in the interstellar medium (ISM) using the
GPU-accelerated {\sc Pencil Code} on the exascale resources of the LUMI
supercomputer\footnote{https://www.lumi.csc.fi/public/}.  We reach higher
magnetic Reynolds numbers than previously possible in the modelling of the
\emph{small-scale dynamo} (SSD) in the ISM.  These results improve our
understanding of how the turbulent magnetic field may be included more
generally in simulations of  galaxy formation and evolution.

%Modelling the growth of galactic magnetic fields requires resolving the
%large separation between driving and turbulent diffusion scales.
%over integration times exceeding 1 Gyr required to evolve the LSD \citep{GEZR08,Gent:2013a, Chamandy16, LKT22}.
%lt With three-quarters of a century of observing
%magnetic fields in the Milky Way and other galaxies \citep{Hiltner1949,
%Hall1949} has revealed the important role they play in galactic formation
%and dynamics. They make an energy contribution similar to that of the kinetic
%turbulence  \citep{KA92, BBMSS96, Beck15}.  At kiloparsec scales, the magnetic
%field affects the multi-phase structure of the ISM by confining the hot gas
%evolving in SN remnants and superbubbles \citep{Hill:2012a, EGSFB19} and by
%reducing turbulent velocities in the ISM \citep{IH17}.  In kiloparsec-scale and
%full galaxy simulations, magnetic fields increase the thickness of the disc,
%extending the halo and reducing its radial extent \citep{PS13, WSPP21, WSPP23,
%Tress24, Bogue25}.
%With respect to star formation in molecular clouds,
%magnetic fields can retard star formation in molecular clouds \citep[e.g.][]%{Mouschovias1999, McKee2007,Crutcher2009, Crutcher2012}
%or enhance it \citep{Zamora18, Sofia21,BRNF25}.
%In full galaxy simulations, magnetic fields can affect the rate andlocation of star formation
%and across galaxies \citep{IH17, Brucy23, Bogue25}. %In particular,
%Cosmic dust, essential to planet and star formation, is shielded by magnetic
%fields from excessive destruction in supersonic blast waves \citep{KMG22}.
%
Modelling the growth of galactic magnetic fields requires resolving the large
separation between driving and turbulent diffusion scales.  Simulations of
large-scale dynamos (LSDs) have yielded solutions with large-scale fields
matching the topology and field strength of those observed but with a turbulent
constituents an order of magnitude weaker \citep{Gressel08b,Gent:2012,GMK24}.
In contrast, observations determine the turbulent field constituent to be as
much as ten times stronger than the mean field \citep{Beck15, BB25}. At the low
resolutions required to have time and resources to evolve the mean field, these
simulations have low magnetic Reynolds numbers, which inhibits the SSD and
weakens the turbulent constituent \citep{Gressel08b,Gent:2012}.  Full galaxy
simulations have even lower Reynolds numbers, especially in the low-density,
high sound speed regions where dynamo action is most important \citep{GMKS22}.
These typically yield merely a turbulent constituent due to insufficient
integration time for the mean-field dynamo to grow \citep{RT16, RT17, RT17a,
Tevlin25}.

%lt \citet{GMK24} find the SSD growth rates to be insensitive to the presence of
%the LSD, but if the LSD is present the turbulent field continues to grow beyond
%the SSD's saturation strength due to tangling.  In contrast, the LSD is slowed
%by the presence of the SSD or strength of the tangling \citep{BSB16,GMK25}. As
%the magnetic Reynolds number increases, the topology of the mean field exhibits
%reversals and switches of parity and its growth decreases.  It is essential for
%understanding the LSD, therefore, to adequately model the SSD and its effects.

\citet{GMKS21} and \citet{GMKS22} find that the magnetic energy at saturation
of the SSD appears to approach an asymptote for magnetic Prandtl numbers $\Pm >
10$, though the magnetic Reynolds number is the determining parameter for the
strength of the field.  However, the grid resolution of those simulations was
restricted to $\dx\in[0.5,4]$~pc.  We, therefore, extend the resolution study
to the range $\dx\in[0.0625,0.5]$~pc.  With any numerical model there is a
contribution to the diffusivity from the numerical scheme, which affects the
effective Reynolds numbers and is difficult to disentangle from the explicit
diffusive parameters in the model. \citet{Kriel22, Kriel25} examine a method
independent of explicit diffusion to determine the effective diffusivity
applying from all sources to a numerical simulation.  Here, however, we simply
identify the explicit Lagrangian magnetic diffusivity $\eta_{\rm crit}$ above
which the solution varies from that with purely numerical diffusion.

%lt In Section \ref{sec:models} we describe our models and in
%Section~\ref{sec:effective} we demonstrate at what resolution we can resolve
%the physical resistivity in our models by testing the dependence of the
%resistivity $\eta$ on resolution over the range $\dx\in[0.5,2]\!\pc$.  In
%Section~\ref{sec:Res} we demonstrate asymptotic behavior of the SSD with
%magnetic Prandtl number and then summarize and conclude in
%Section~\ref{sec:Sum}.

%-------------------------------------------------------------------------o
\section{Model design}\label{sec:models}
%--------------------------------------------------------------------------
%-------------------------------------------------------------------------
\begin{table}
\caption{
Models
\label{tab:models}
}
\begin{center}
\begin{tabular}{lcccc}
\hline\hline
Model &$\dx$ &  Pm                   &$\nu_6,\chi_6,\eta_6$ &$R_{\rm min}$  \\
      &[pc]  &                       &[kpc$^5\kms$]        &[pc]   \\\hline
\NAA  &2     &[\quad1,$\infty$\quad] &  3 (-14)             &7.0    \\
\NBA  &1     &[\quad1,$\infty$\quad] &  1 (-15)             &5.0    \\
\NCA  &1/2   &[\quad1,$\infty$\quad] &  4 (-17)             &4.5    \\
\PCB  &1/2   &[\quad1,20\quad]       &  4 (-17)             &4.5    \\
\PDC  &1/4   &[\quad1,100~\,]        &  2 (-18)             &3.8    \\
\PEE  &1/8   &[\,~10,500~\,]         &  8 (-20)             &3.4   \\\hline
%\PFG  &1/16  &[100,1000]             &  4 (-21)             &3.2   \\\hline
\end{tabular}
\end{center}
\tablecomments{Models to benchmark dependence of
$\eta_{\rm crit}$ on resolution are indicated by N and for testing sensitivity
to Pm by P, with the number of grid points on a side given.  Viscosity
$\nu=10^{-3}\kpc \kms$ with the range of $\Pm=\nu/\eta$ given in square brackets.
Sixth order hyperdiffusion coefficients are $\nu_6=\chi_6=\eta_6$.  $R_{\rm
min}$ is the minimum radial scale $R_{\rm SN}$ of the SN energy injection
profile. Parenthesized values give powers of ten.}
\end{table}
%--------------------------------------------------------------------------
Our model domain is a periodic cubic domain of size $h=256\pc$ on each side
containing monatomic gas with  mean number density $1\cmcube$. SNe explode
intermittently at a Poisson rate $\dot\sigma=0.75\SNr$ at uniform random
locations, where the solar neighbourhood rate $\SNr\simeq 17\kpc^{-2}\Myr^{-1}$
\citep{BT91,Mannucci05}. The volume SN rate is determined using a galactic
thickness, which we choose to be the side $h$ of our periodic box. The periodic
domain excludes pressure release by galactic winds or fountains, so we choose
an SN rate lower than the solar neighborhood value to avoid thermal runaway
\citep{KO15} and maintain a multiphase ISM for the duration of the simulations.

Our models are summarised in Table\,\ref{tab:models}.  Those denoted by N apply
the kinematic phase of the SSD for benchmarking the magnetic diffusivity
$\eta_{\rm crit}$ that can be resolved at each resolution
(Sect.\,\ref{sec:effective}).  Those denoted by P evolve the SSD for
$\eta>\eta_{\rm crit}$ until saturation of the dynamo is established
(Sect.\,\ref{sec:Res}). As a common initial condition for the set of P models,
a snapshot from model \NCA\ during its kinematic phase is used. For models with
higher resolution, it is remeshed employing cubic interpolation.

To run our models, we use the {\sc Pencil Code}\footnote{\href{https://pencil-code.nordita.org/}{\rev{https://pencil-code.nordita.org/}}} \citep{Pencil-JOSS} coupled
with the {\sc Astaroth}\footnote{\href{https://bitbucket.org/jpekkila/astaroth/src/master/}{\rev{https://bitbucket.org/jpekkila/astaroth/src/master/}}} library \citep{Pekkila19, V21, Pekkila22}. This allows us to
use GPUs for speedup by as much as a factor of 20 for this problem.  We solve
the compressible, non-ideal, non-adiabatic system of magnetohydrodynamic
equations
  \begin{eqnarray}
  \label{eq:mass}
    \frac{\text{D}\rho}{\text{D}t} &=&
    -\rho \nabla \cdot \vect{u}
    +\nabla \cdot\zeta_D\nabla\rho,%\\[0.3cm]
  \end{eqnarray}\vspace{-0.5cm}
  \begin{eqnarray}
  \label{eq:mom}
    \rho\frac{\text{D}\vect{u}}{\text{D}t} &=&\mu_0^{-1}\nabla\times\vect{B}\times\vect{B}
    -\rho\cs^2\nabla\left({s}/{\cplocal}+\ln\rho\right)\nonumber\\
    &+&
    \nabla\cdot \left(2\rho\nu{\mathbfss W}\right)
    +\rho\nabla\left(\zeta_{\nu}\nabla \cdot \vect{u} \right)
    \nonumber\\
    &+&\nabla\cdot \left(2\rho\nu_6{\mathbfss W}^{(5)}\right)
  -\vect u\nabla\cdot\left(\zeta_D\nabla\rho\right),%\\[0.3cm]
  \end{eqnarray}\vspace{-0.5cm}
  \begin{eqnarray}
  \label{eq:ind}
    \frac{\partial \vect{A}}{\partial t} &=&
    \vect{u}\times\vect{B}
    +\eta\nabla^2\vect{A}
    +\eta_6\widetilde\nabla^6\vect{A},%\\[0.3cm]
  \end{eqnarray}\vspace{-0.5cm}
  \begin{eqnarray}
  \label{eq:ent}
    \rho T\frac{\text{D} s}{\text{D}t} &=&
    \EST\dot\sigma h^{-1} +\rho\Gamma
    -\rho^2\Lambda +\eta\mu_0^{-1}|\nabla\times\vect{B}|^2
    \nonumber\\
    &+&2 \rho \nu\left|{\mathbfss W}\right|^{2}
    +\rho~\zeta_{\nu}\left(\nabla \cdot \vect{u} \right)^2
    \nonumber\\
    &+&\nabla\cdot\left(\zeta_\chi\rho T\nabla s\right)
    +\rho T\chi_6\widetilde\nabla^6 s
    \nonumber\\
    &-& \cv\gamma~T \left(
    \zeta_D\nabla^2\rho + \nabla\zeta_D\cdot\nabla\rho\right),
  \end{eqnarray}
where $\rho$ denotes the density of the gas, $s$ its specific entropy, $T$ its
temperature, and $\vect{u}$ its velocity, with $\cs$ the adiabatic sound speed,
$t$ the time; $\vect{A}$ is the magnetic vector potential, with magnetic field
$\vect{B}=\nabla\times\vect{A}$. The constant $\mu_0$ denotes the vacuum
magnetic permeability and $\cplocal$ and $\cv$ the gas specific heat capacity
at constant pressure and constant volume, respectively. The advective
derivative
\begin{equation}\label{eq:adv}
\frac{\text{D}~}{\text{D}t}=\frac{\partial~}{\partial t} + \vect{u}\cdot\nabla,
\end{equation}
 the traceless rate of strain tensor
 $\mathbfss W$ has components
\begin{equation}
 {\mathsf W}_{ij} = \dfrac{1}{2}\left(\dfrac{\partial u_i}{\partial x_j}
                  + \dfrac{\partial u_j}{\partial x_i}
                  -\dfrac{2}{3} \delta_{ij}\nabla\cdot \vect u\right),
\end{equation}
and $|\mathbfss W|^2={\mathsf W}_{ij}^2$ is its contraction.\footnote{We adopt
Einstein's summation convention throughout.}

To manage numerical instabilities near the Nyquist frequency, we apply a sixth
order hyperdiffusivity in Eq.\,\eqref{eq:mom}, proportional to the divergence
of $\rho {\mathbfss W}^{(5)}$, where
\begin{equation}
 {\mathsf W}_{ij}^{(5)} = \widetilde\nabla^4  {\mathsf W}_{ij},
\end{equation}
and in Eqs.\,\eqref{eq:ind} and \eqref{eq:ent} proportional to
$\widetilde\nabla^6\vect{A}$ and $\widetilde\nabla^6 s$, respectively
with\footnote{Distinct from the biharmonic operator
\\$\nabla^4=(\partial^2/\partial x_i^2)^2$, etc.}
\begin{equation}
\widetilde\nabla^4 =\dfrac{\partial^2}{\partial x_i^2}\dfrac{\partial^2}{\partial x_i^2}, \quad
\widetilde\nabla^6 =\dfrac{\partial^3}{\partial x_i^3}\dfrac{\partial^3}{\partial x_i^3}.
\end{equation}
The corresponding coefficients $\nu_6=\eta_6=\chi_6$ are listed in
Table\,\ref{tab:models}.  Models at each resolution were tested to find the
minimum of $\nu_6$, for which the corresponding grid Reynolds number at the
Nyquist frequency
\begin{equation}
\Rey_{}^{(6)}=\dfrac{u_{\rm rms}}{\nu_6k_{\rm Ny}^5}\;,
\end{equation}
typically peaking shortly after each explosion, would remain below 10
\citep[see][which gives more detail on the use of hyperdiffusivity]{HB04}.

To resolve shock discontinuities, artificial diffusion with coefficients
\begin{eqnarray}
\begin{matrix}
\zeta_D,\\ \zeta_\nu,\\ \zeta_\chi\
\end{matrix}
=
\begin{cases}
0.3\,|\nabla\cdot\vect{u}|^{7/4}\kpc\kms,   & \!\!\nabla\cdot\vect{u}<0\\\nonumber
0.015\,|\nabla\cdot\vect{u}|^{7/4}\kpc\kms, & \!\!\nabla\cdot\vect{u}>0
\end{cases}
\end{eqnarray}
is applied in Eqs.\,\eqref{eq:mass}, \eqref{eq:mom}, and \eqref{eq:ent},
respectively.  The terms with $\zeta_D$ and $\zeta_\chi$ are almost quadratic
in $\nabla\cdot\vect{u}$, while the one with $\zeta_\nu$ is almost cubic.
Commonly, such a scheme has been applied only to convergent flows and with the
exponent 1 instead of $7/4$ \citep[e.g.,][with prefactor 1 instead of
0.3]{GMKSH20}. We find it necessary to apply it also, albeit more weakly, to
divergent flows to suppress numerical instability there, while the higher
exponent better confines the diffusion locally to the shocks.

Expressions containing $\zeta_D$ are also included in Eqs.\,\eqref{eq:mom} and
\eqref{eq:ent} to correct for momentum and energy conservation. A missing
factor in our earlier work of $\gamma=\cplocal/\cv$ has been included in
Eq.~\eqref{eq:ent}.  \citet{Sankalp25} found that this correction term
prefactor in Eq.~\eqref{eq:ent} significantly impacted the thickness of their
modelled solar atmosphere, but it has limited effect in the unstratified
turbulent ISM modelled here.

The system is driven and regulated by sources and sinks of heat in
Eq.\,\eqref{eq:ent}. SNe are applied with $\EST=10^{51}\erg$ injected in
spheres with a Gaussian radial thermal energy profile of scale $R_{\rm SN} \geq
R_{\rm min}\in[3.1875,7]\pc$ (as per $\dx$, see Table\,\ref{tab:models}),
always chosen to be large enough to contain at least $100\,M_\odot$ of ambient
gas within $R_{\rm SN}$.  Neither mass nor momentum are included in the SN
injection.
%lt that follows the functional form given by \citet{Wolfire:1995} with temperature dependence approximated as
%\begin{equation}\label{eq:Gamma}
%\Gamma(T) = \frac{\Gamma_0}{2}\left(1+\tanh\left[\frac{
%     2\cdot 10^4\,{\rm K} - T}{2000\,{\rm K}}\right]\right)\!,
%\end{equation}
%and $\Gamma_0 = 0.0147$\,erg\,g$^{-1}$\,s$^{-1}$.  The cooling applies a
%piecewise power-law temperature dependence $\Lambda (T) = \Lambda_k
%T^{\beta_k}$ within the range $T_k < T < T_{k+1}$ based on \citet{Wolfire:1995}
%for cold and warm gas phases and \citet{Sarazin:1987} for the hot phase, as
%parameterized previously \citep[e.g,][]{SVG02,Gressel:2008}, see
Cooling following \citet{SVG02} and \citet{Gressel:2008} is implemented
\citep[see][Table 1]{Gent:2013b} but with $\Lambda=0$ for $T<90$~K, which
produces thermally unstable ranges at high and low temperatures as expected for
the ISM. Moreover, uniform far ultraviolet heating $\Gamma(T)$ is included
\citep{Wolfire:1995}.
%\citep{Gent:2013b}.

The ideal gas equation of state closes the system, assuming an adiabatic index
$\gamma =\cplocal/\cv=5/3$.  Treating the ISM as a monatomic, fully ionized
plasma we apply a mean molecular weight of 0.531. An important revision to our
prior implementations is to replace third-order time integration, constrained
primarily by Courant conditions, with a fourth-order accuracy-constrained
adaptive time stepping scheme following \citet{Kennedy00}.
%--------------------------------------------------------------------------
\section{Resolution of magnetic diffusivity}\label{sec:effective}
%--------------------------------------------------------------------------
We choose a viscosity $\nu=10^{-3}\kpc\kms$ because we have demonstrated that
this value exceeds numerical viscosity at any resolution used in this study
\citep{GMKS21, GMKS22}. By this, we mean that we can distinguish a difference
in the energy spectrum from the solution without physical viscosity.

\begin{figure}
\centering
\includegraphics[trim=0.2cm 1.6cm 0.0cm 0.1cm,clip=true,width=\columnwidth]{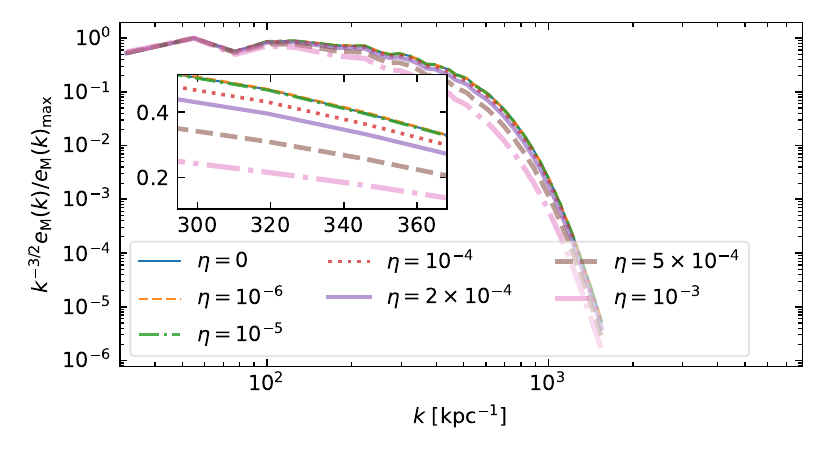}
\includegraphics[trim=0.2cm 1.6cm 0.0cm 0.1cm,clip=true,width=\columnwidth]{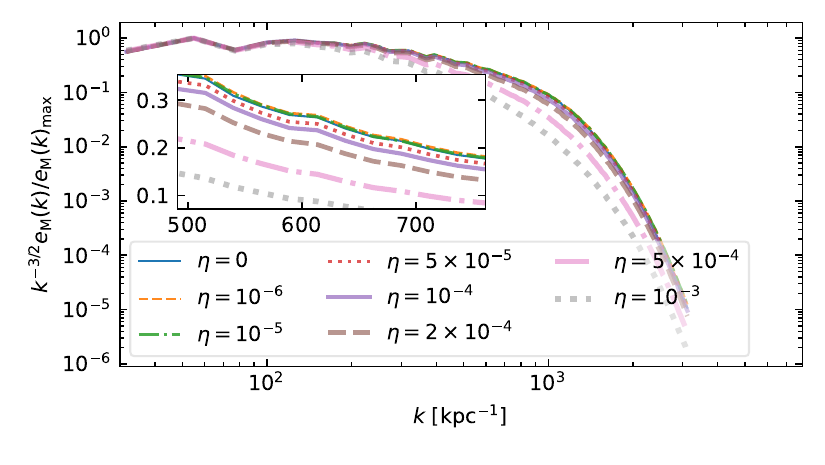}
\includegraphics[trim=0.2cm 0.5cm 0.0cm 0.1cm,clip=true,width=\columnwidth]{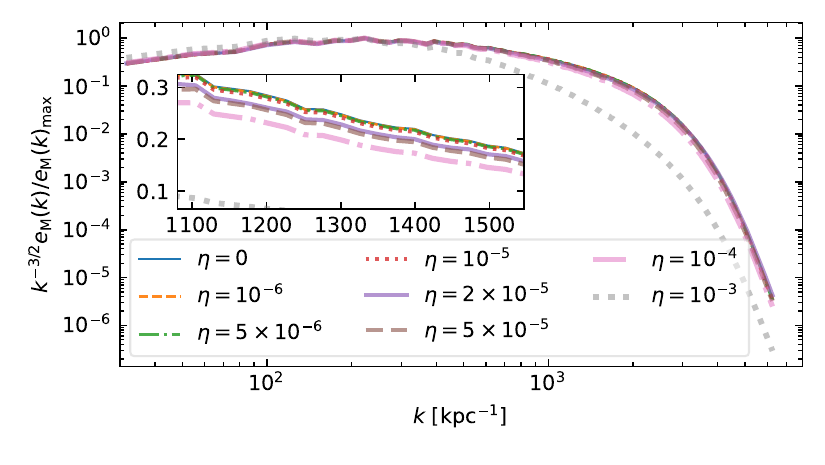}
\includegraphics[trim=0.1cm 0.5cm 0.1cm 0.3cm,clip=true,width=\columnwidth]{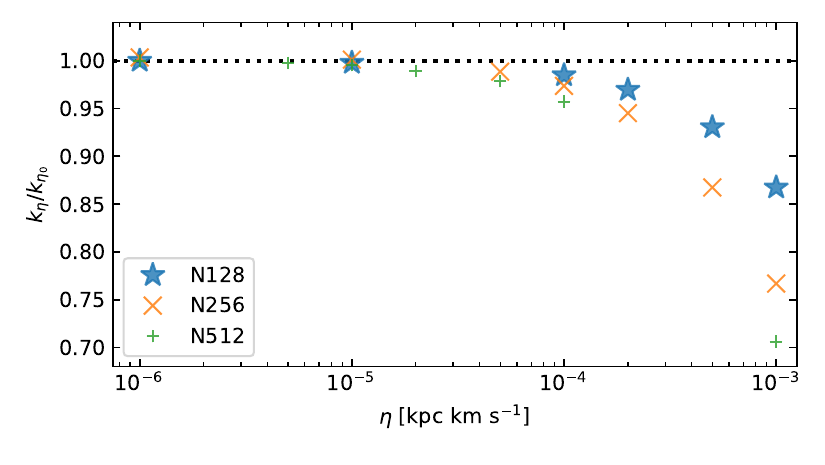}
\caption{
Magnetic energy spectra $e_{\rm \rev{M}}(k)$ compensated by the Kazantsev
scaling $k^{3/2}$ and normalized by their compensated maximum for \emph{(a)}
models \NAA, \emph{(b)} \NBA\ and \emph{(c)} \NCA\ using different values of
the magnetic diffusivity $\eta$ in$\kpc\kms$ as listed in the legends.  Insets
zoom in near the start of the inertial range to highlight deviations from the
curve for $\eta=0$.  \rev{In panel \emph{(d)},  the wavenumber
$k_\eta$ at the dissipation scale, normalised by $k_{\eta_0} =
k_{\eta}(\eta=0)$, is plotted against $\eta$. The dissipation scale is resolved, defining $\eta_\text{crit}$,
for values of $\eta$ where $k_\eta/k_{\eta_0} < 1$.}
\label{fig:eta-norm}
}
 \begin{picture}(0,0)
    \put(-122,646){{\sf\bf{(a)}}}
    \put(-122,535){{\sf\bf{(b)}}}
    \put(-122,424){{\sf\bf{(c)}}}
    \put(-122,298){\rev{{\sf\bf{(d)}}}}
  \end{picture}
\end{figure}
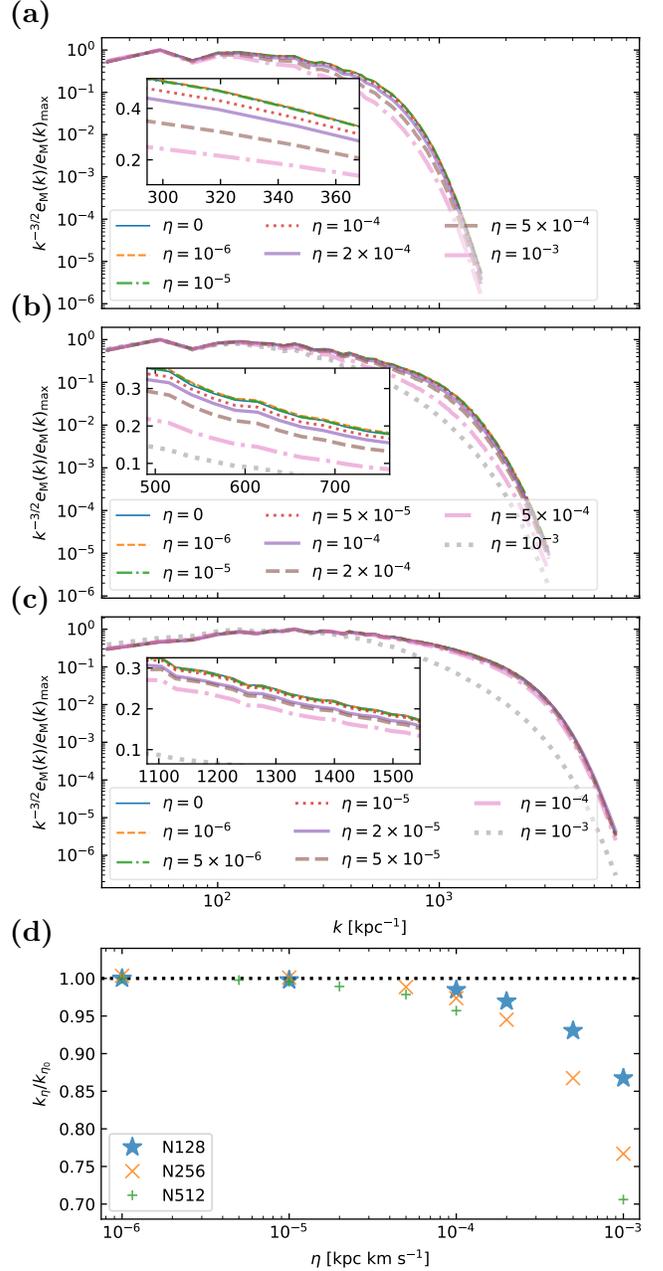

To identify asymptotic behavior of the SSD saturation level with respect to the
magnetic Reynolds number $\Rm=\lf u_{\rm rms}/\eta$ or Prandtl number
Pm$\,=\nu/\eta$, we need to choose explicit magnetic diffusivities $\eta$ in
our simulations that exceed the numerical diffusivity.  Hence, we test what
values of $\eta$ distinguish the solution from that with only hyperdiffusion
and shock diffusion in runs during the kinematic phase of the SSD, labelled
with N in Table\,\ref{tab:models}.  The magnetic energy spectra from these
runs, compensated by the Kazantsev $k^{3/2}$ scaling, are plotted in
Figure~\ref{fig:eta-norm}. The insets zoom in near the start of the dissipation
range, to allow identification of the critical diffusivity $\eta_{\rm crit}$
above which the deviations from the solutions for $\eta=0$ become significant.
These are listed in Table\,\ref{tab:crit} and indicate that $\eta_{\rm crit}
\approx 5\times 10^{-5} (\dx / 1 \mbox{ pc}) \kpc\kms$.

\rev{This is confirmed quantitatively by estimating the wavenumber at the
dissipation scale \citep[see Section 12.1,][given $k_\eta=2\uppi/l_\eta$]{MY07}
\begin{equation}\label{eq:k2}
  k_\eta =\left(\dfrac{\sum_k k^2 e_\text{M}(k)}{\sum_k e_\text{M}(k)}\right)^{1/2}.
\end{equation}
To mitigate statistical noise, the energy spectra are aggregated over the final
24 snapshots for models \NAA--\NCA, spanning 1.2~Myr. The dissipation scales for models without explicit diffusivity $k_{\eta_0}=k_\eta(\eta=0)$ are listed in
Table~\ref{tab:crit}.  In Figure~\ref{fig:eta-norm}\emph{(d)},  $k_\eta$
normalized by $k_{\eta_0}$ deviates from unity at $\eta>\eta_\text{crit}$.
}

%------------------------------------------------------------------------
\begin{table}
\caption{
Critical magnetic diffusivity
\label{tab:crit}}
\begin{center}
\begin{tabular}{cccc}
\hline\hline
Model &$\dx$ &  $\eta_{\rm crit}$ &\rev{$k_{\eta_0}$}      \\
      &[pc]  &[kpc$\kms$]         &\rev{[kpc$^{-1}$]     } \\\hline
\NAA  &2     &  1 (-4)            &\rev{  \phantom{0}460}  \\
\NBA  &1     &  5 (-5)            &\rev{  \phantom{0}875}  \\
\NCA  &1/2   &  2 (-5)            &\rev{  1772          }  \\
\hline
\end{tabular}
\end{center}
\tablecomments{Minimum magnetic diffusivity $\eta_{\rm crit}$ required to
modify the solution at each resolution. Parenthesized values give powers of
ten.  \rev{The dissipation wavenumber $k_{\eta_0}$ for the magnetic energy at $\eta=0$,
 is obtained with Eq.~\eqref{eq:k2}.}
}
\end{table}
%------------------------------------------------------------------------

%-------------------------------------------------------------------------
%--------------------------------------------------------------------------

%-------------------------------------------------------------------------
\begin{figure*}
\centering
\includegraphics[trim=0.2cm 1.6cm 0.0cm 0.1cm,clip=true,width=\columnwidth]{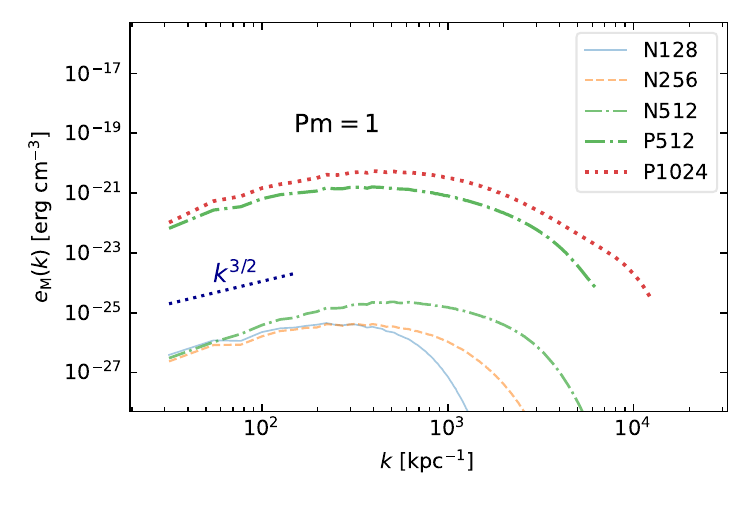} \includegraphics[trim=0.2cm 1.6cm 0.0cm 0.1cm,clip=true,width=\columnwidth]{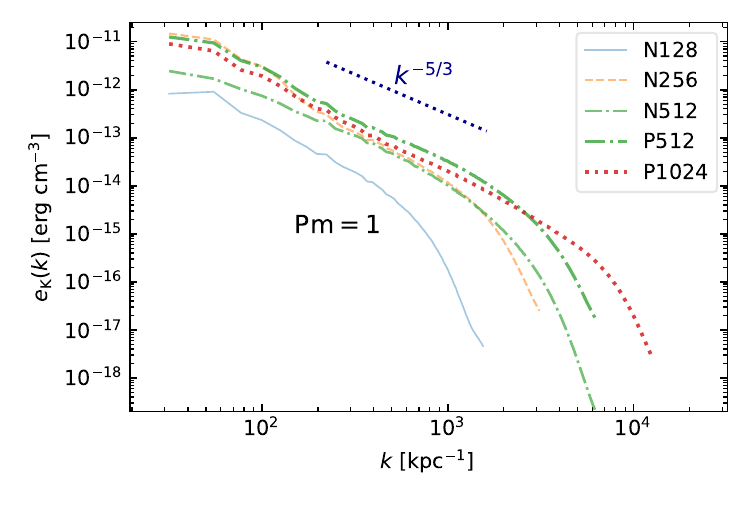}
\includegraphics[trim=0.2cm 1.6cm 0.0cm 0.1cm,clip=true,width=\columnwidth]{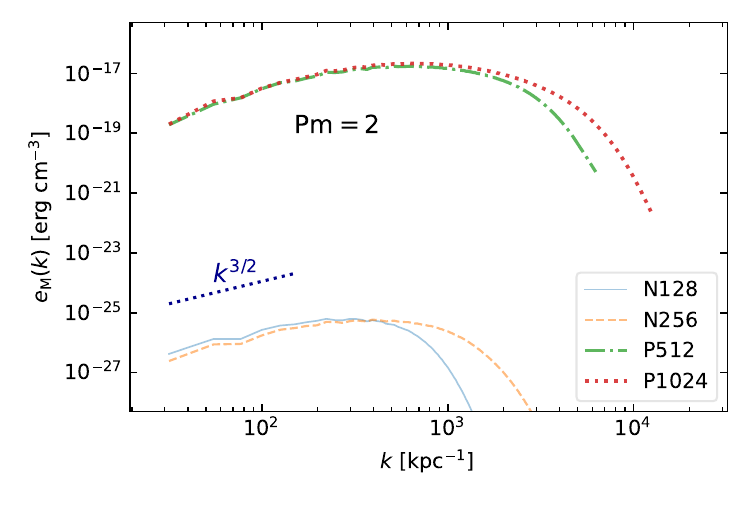} \includegraphics[trim=0.2cm 1.6cm 0.0cm 0.1cm,clip=true,width=\columnwidth]{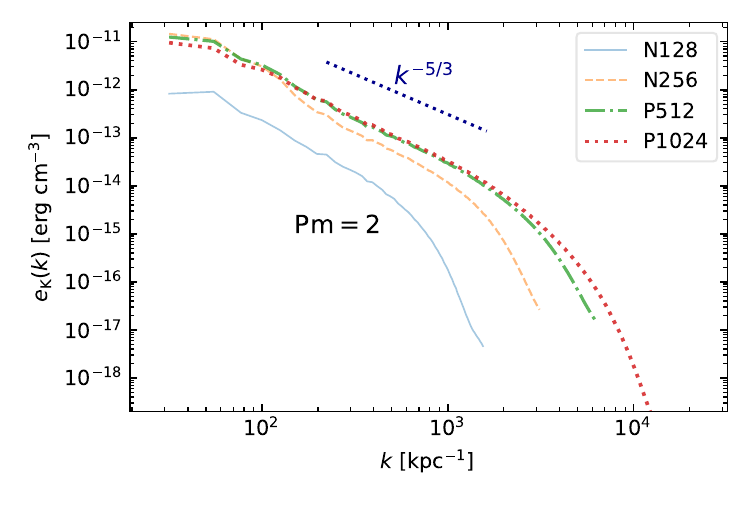}
\includegraphics[trim=0.2cm 1.6cm 0.0cm 0.1cm,clip=true,width=\columnwidth]{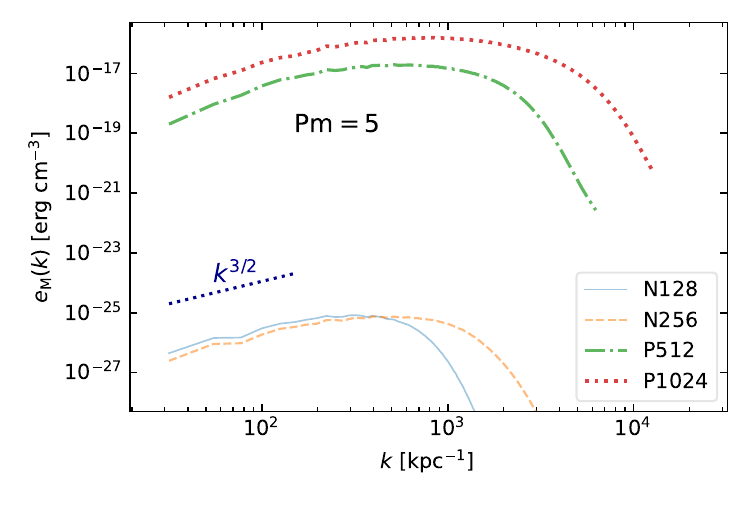} \includegraphics[trim=0.2cm 1.6cm 0.0cm 0.1cm,clip=true,width=\columnwidth]{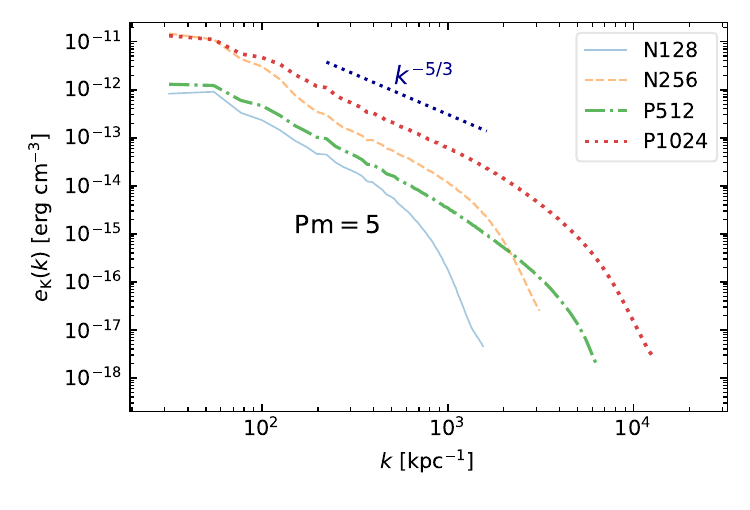}
\hspace*{+0.1mm}\includegraphics[trim=0.2cm 0.2cm 0.0cm 0.1cm,clip=true,width=\columnwidth]{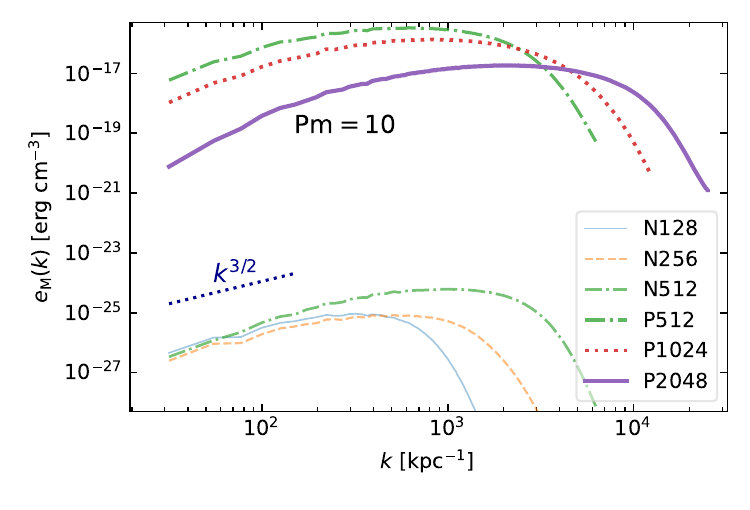}\hspace*{1.1mm}\includegraphics[trim=0.2cm 0.2cm 0.0cm 0.1cm,clip=true,width=\columnwidth]{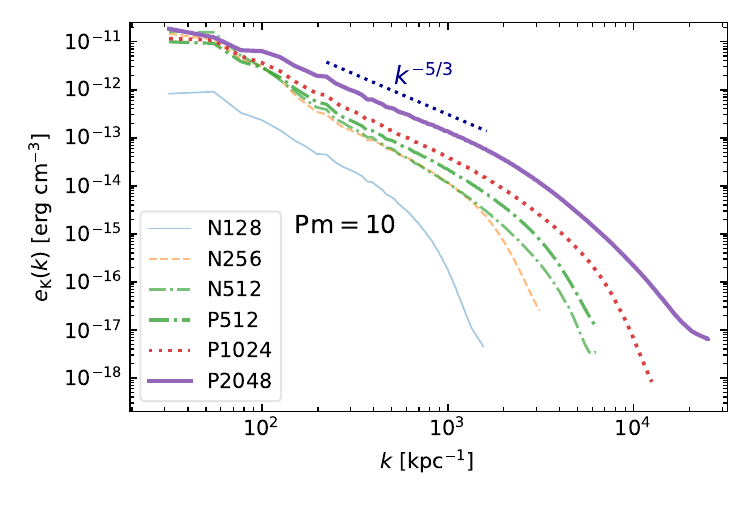}
\caption{
%FAG original normalised
%Magnetic energy spectra $e_{\rm \rev{M}}(k)$ \rev{\emph{(a)--(d)}} normalized by $e_{\rm \rev{M}}(k_f)$ at a
%typical SN forcing wavenumber $k_f = 125 \mbox{ kpc}^{-1}$
% The $k^{3/2}$ Kazantsev scaling is indicated
%upper left (dotted dark blue).
%FAG: alt with energy density
\rev{\emph{(a)--(d)} Magnetic energy spectra $e_{\rm \rev{M}}(k)$}
at a given Pm for models listed in the legends.
 The $k^{3/2}$ Kazantsev scaling is indicated
\rev{lower} left \emph{(dotted dark blue line)}.
\rev{\emph{(e)--(h)} Corresponding kinetic energy spectra $e_{\rm K}(k)$. The $k^{-5/3}$ Kolmogorov scaling is indicated in the upper centre \emph{(dotted dark blue line)}.}
\label{fig:etak-res}
}
 \begin{picture}(0,0)
    \put(-238,618){{\sf\bf{(a)}}}
    \put(-238,481){{\sf\bf{(b)}}}
    \put(-238,347){{\sf\bf{(c)}}}
    \put(-238,212){{\sf\bf{(d)}}}
    \put(   2,618){{\sf\bf{(e)}}}
    \put(   2,481){{\sf\bf{(f)}}}
    \put(   2,347){{\sf\bf{(g)}}}
    \put(   2,212){{\sf\bf{(h)}}}
  \end{picture}
\end{figure*}
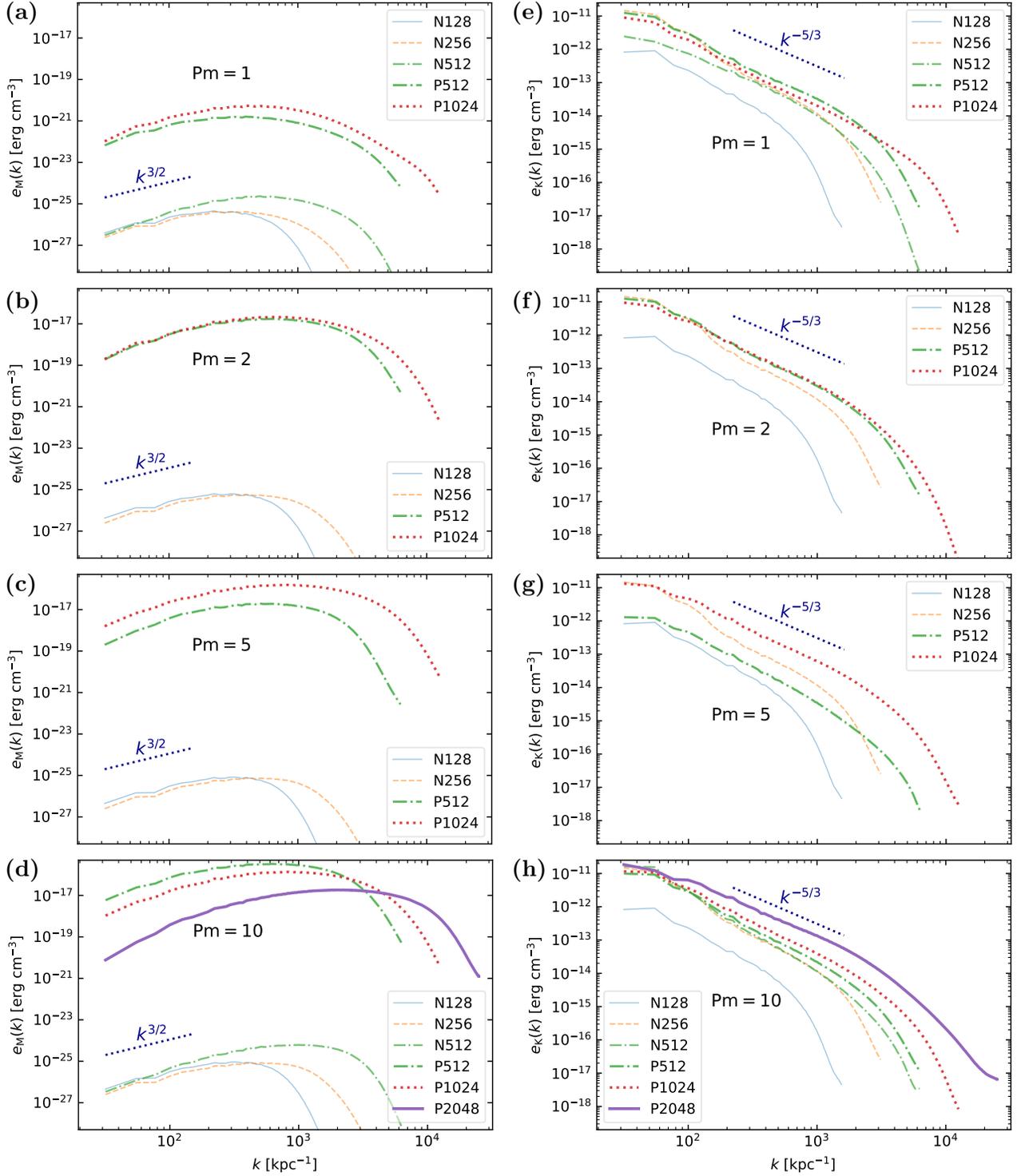
%--------------------------------------------------------------------------
%--------------------------------------------------------------------------
\section{Results}\label{sec:Res}
%--------------------------------------------------------------------------
\subsection{Spectral scaling}
Magnetic energy spectra for various Pm are displayed in
Figure~\ref{fig:etak-res} for both N and P model sets.
%FAG: Revised without k_f normalisation, para break removed
%FAG: Alternate text for original plots with k_f  normalisation
%To ease comparison
%between results during varying stages of the kinematic phase up to and
%including saturation, each spectrum is normalized by its magnetic energy at the
%forcing scale, $e_{\rm \rev{M}}(k_f)$, where $l_f=2\uppi k_f^{-1}\simeq50\pc$ is the
%typical forcing scale of the SN remnants as determined by the correlation
%length of the turbulent velocity near the galactic midplane \citep{HSSFG17}.
%
%FAG: alternate
In Figure~\ref{fig:etak-res}\emph{(a)} with $\Pm=1$, the spectra for the models
of set N in the kinematic phase align with the Kazantzev $k^{3/2}$ scaling for
$k<\kf$ \citep{K68} while the peak energy extends to higher $k$ with increasing
resolution\rev{, where the forcing wavenumber $\kf=2\uppi \lf^{-1}$
 is
determined by the correlation
length of the turbulent velocity near the galactic midplane $\lf\simeq50\pc$ \citep{HSSFG17}.}

The Kazantsev scaling is also present for the models~\PCB\ and
\PDC, but relative to model~\NCA, the energy peak is pushed to lower $k$,
reflecting the apparent saturation of the SSD. This is more clearly evident in
Figure~\ref{fig:etak-res}\emph{(d)} for $\Pm=10$ when comparing the spectra for
model~\PCB\ with \NCA\ (green, dash-dotted, thick and thin, respectively).
Model~\PCB\ has reached saturation and its energy peak is at significantly
lower $k\simeq700\kpc^{-1}$ than for model~\NCA\ at $k\simeq1300\kpc^{-1}$.
Model~\PDC\ is also approaching saturation with peak energy at  even lower $k$
than in model~\NCA, while model~\PEE\ is still kinematic with its peak located
at even higher $k$.  In summary, the peak energy in saturated fields converges
at wavenumbers far below the Nyquist wavenumber, and Kazantzev scaling holds at
lower wavenumbers.

\rev{In Figure~\ref{fig:etak-res}\emph{(e)--(h)}, the kinetic energy at low resolution \NAA\ is damped for all $k$, because numerical diffusion has effects beyond the dissipation range \citep[see][which shows that this can lead to false convergence]{GMKS21}.
Above this resolution, the spectra remain well converged.
The inertial range is slightly steeper than
$k^{-5/3}$. Comparing \NCA\ with \PCB\ for $\Pm=10$, the
kinetic energy is enhanced preferentially
at $k>k_\eta$ in the nonlinear SSD.}

\subsection{Saturation}

We study the saturation of models with varying Pm, Rm, and numerical
resolution, always ensuring $\eta > \eta_{\rm crit}$, so the physical magnetic
diffusivity determines the solution.

\begin{figure}
\centering
\includegraphics[trim=0.4cm 0.35cm 0.02cm 0.375cm,clip=true,width=0.95\columnwidth]{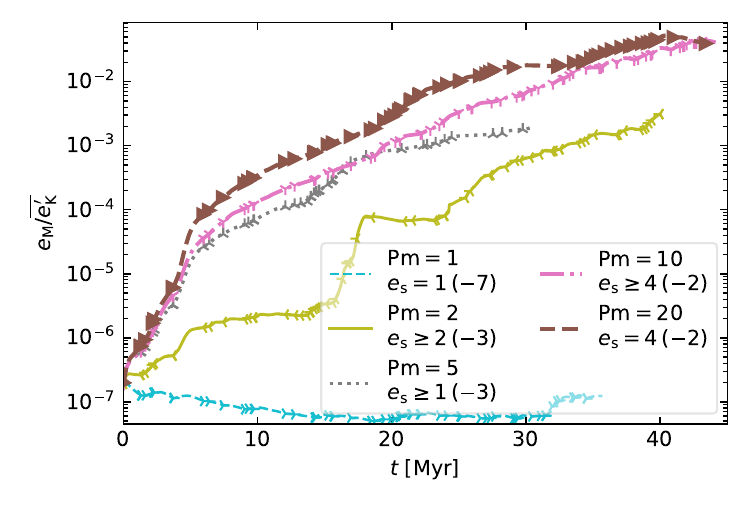}
\includegraphics[trim=0.4cm 0.35cm 0.25cm 0.375cm,clip=true,width=0.95\columnwidth]{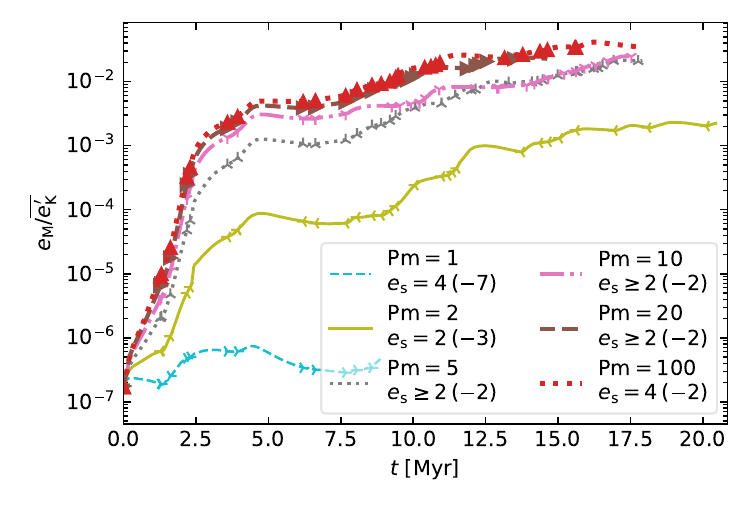}
\includegraphics[trim=0.4cm 0.35cm 0.10cm 0.375cm,clip=true,width=0.95\columnwidth]{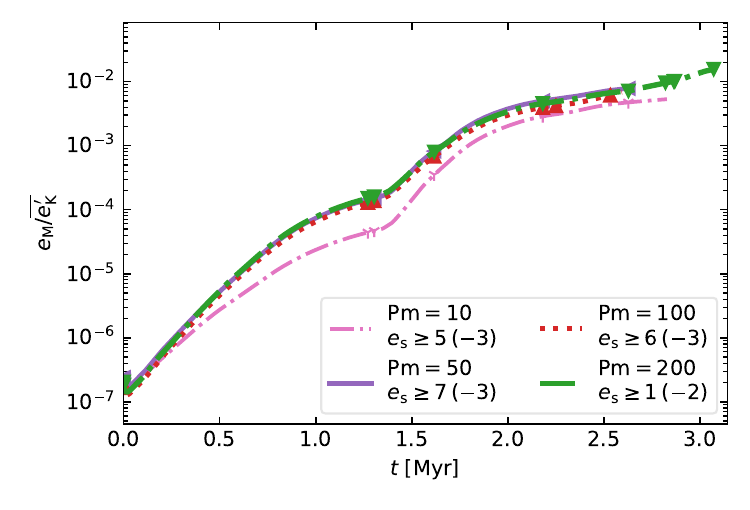}
\caption{
Magnetic energy $e_{\rm \rev{M}}$ normalised by time-averaged turbulent kinetic
energy $\eKt$ for grid \rev{sizes} \emph{(a)} $512^3$\rev{,} \emph{(b)} $1024^3$\rev{, and} \emph{(c)}
$2048^3$~\rev{ zones.}
See the legends for Pm and \rev{saturation energy} $e_{\rm s}$ from
an average over the \rev{final} 2.5\% of simulation time. Markers are shown at the
time of SN explosions.  Note that the time scales are dramatically smaller at
higher resolution.
\label{fig:eBt-Pm}
}
 \begin{picture}(0,0)
    \put(-117,562){{\sf\bf{(a)}}}
    \put(-117,409){{\sf\bf{(b)}}}
    \put(-117,257){{\sf\bf{(c)}}}
  \end{picture}
\end{figure}
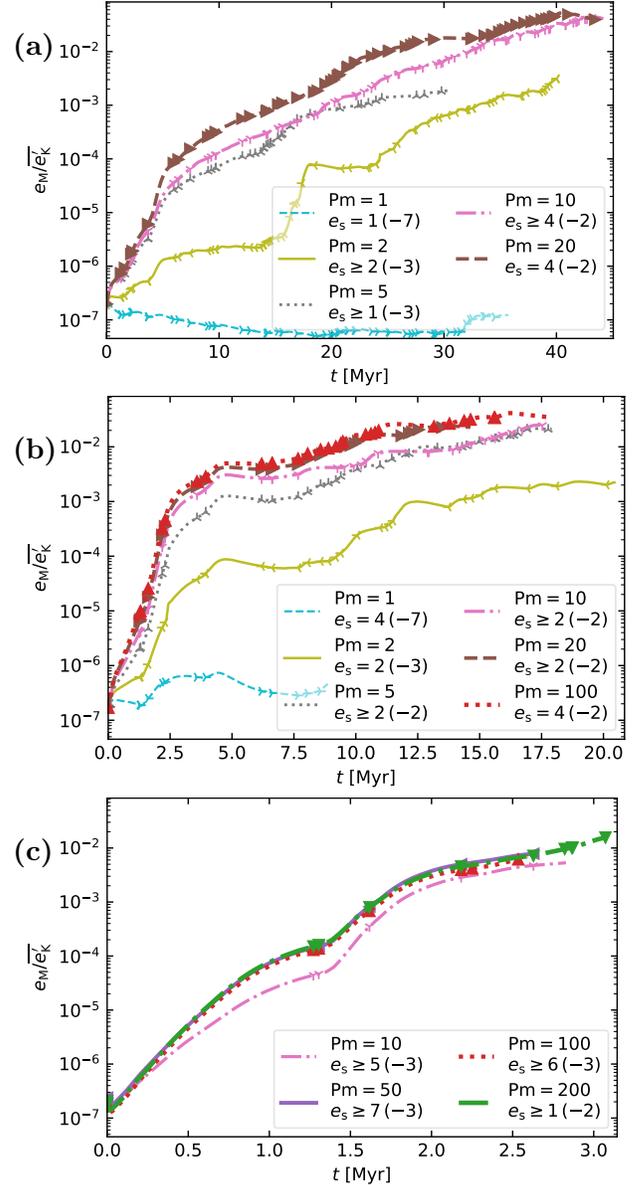

Figure~\ref{fig:eBt-Pm} shows the evolution of the magnetic energy density
$e_{\rm \rev{M}}$ normalized by the time-averaged kinetic energy density
$\overline{e_{\rm K}}$ for each simulation.  Although the only parameter
varying at fixed resolution is $\eta$, the models can diverge over time as the
system is inherently chaotic and the time step differs in each realisation. We
therefore mark SN explosions, which can be followed by sudden surges of
magnetic energy growth.
%-------------------------------------------------------------------------
%--------------------------------------------------------------------------

Note that growth is substantially faster at higher resolution \citep{GMKS21},
and the time axis covers a correspondingly shorter period of time in each
panel.  This indicates that the properties of the turbulence are not completely
determined by the explicit values of $\nu$ and $\eta$ but also by the higher
numerical diffusion at lower resolution.

We measure the strength of the magnetic energy after the SSD has saturated
$e_{\rm s}$ relative to $\overline{e_{\rm K}}$ during the final 2.5\% of run
time for each  simulation, and list these in the legends of
Figure~\ref{fig:eBt-Pm} and in Table\,\ref{tab:Rm}.  Due to our finite
numerical resources, the higher resolution models in
Figure~\ref{fig:eBt-Pm}\emph{(c)} and \emph{(d)} are not yet saturated\rev{, so}
$e_{\rm s}$ is a lower limit to the saturation level in these models.  Some of
the models in Figure~\ref{fig:eBt-Pm}\emph{(a)} and \emph{(b)} also do not
appear to have yet fully saturated, so it would be preferable to extend those
runs as well.  For resolution $N=2048^3$ in Figure~\ref{fig:eBt-Pm}\emph{(c)},
$e_{\rm s}=0.02$ for $\Pm=10$ is consistent with levels obtained at the lower
resolution.  At higher Rm, $e_{\rm s}$ remains below those of matching Rm runs
at lower resolution.  The solutions are close in Figure~\ref{fig:eBt-Pm} at
$\Pm=20$ for models~\PCB\ and \PDC\ and the spectra are converging for these
two models at $\Pm=10$ in Figure~\ref{fig:etak-res}\emph{(d)}.  We find that
the saturation level of the magnetic field approaches a plateau at around 5\%
of equipartition with the kinetic energy independent of resolution as $\Pm$
increases above $10$.

%------------------------------------------------------------------------
\begin{table*}
\caption{
SSD saturation energy
\label{tab:Rm}}
\begin{center}
\begin{tabular}{@{\hspace{-0mm}}c@{\hspace{1.5mm}}r@{\hspace{4.5mm}}rrr@{\hspace{6.5mm}}lll@{\hspace{6.5mm}}lllc}% c}
\hline\hline\\[-4mm]
$\eta$          &Pm  & \multicolumn{3}{c}{Rm}                 & \multicolumn{3}{c}{$e_{\rm s}$}        & \multicolumn{3}{c}{\rev{$L_B$}}                               &\\
$[\!\!\kpc\kms]$&    & \multicolumn{3}{c}{[$10^3$]}           & \multicolumn{3}{c}{$[\eKt]$}           & \multicolumn{3}{c}{\rev{[pc]}}                                &\\
                &    & \multicolumn{3}{c}{at resolution [pc]} & \multicolumn{3}{c}{at resolution [pc]} & \multicolumn{3}{c}{\rev{at resolution [pc]}}                  &\\
                &    & $1/2$  & $1/4$  & $1/8$                & $1/2$  & $1/4$  & $1/8$                & \rev{$1/2$}           &\rev{$1/4$}           & \rev{$1/8$}    &\\
\hline
1 (-3)          &1   &2.8     &  3.0   &                      & 1 (-7)*& 4 (-7)*&                      & \rev{11.3*}           &\rev{10.1*}&                           &\\
5 (-4)          &2   &5.8     &  6.3   &                      & 2 (-3) & 2 (-3)*&                      & \rev{\phantom{0}8.8 } &\rev{\phantom{0}8.5*} &                &\\
2 (-4)          &5   &9.1     &   20   &                      & 1 (-3) & 2 (-3) &                      & \rev{11.0 }           &\rev{\phantom{0}6.4 } &                &\\
1 (-4)          &10  &26      &   29   &  53                  & 4 (-2) & 2 (-2) & 5 (-5)               & \rev{\phantom{0}9.4 } &\rev{\phantom{0}6.0 } & \rev{2.6}      &\\
5 (-5)          &20  &47      &   54   &                      & 4 (-2)*& 2 (-2) &                      & \rev{11.5*}           &\rev{\phantom{0}6.2 } &                &\\
2 (-5)          &50  &        &        & 263                  &        &        & 7 (-3)               &                       &                      & \rev{2.4}      &\\
1 (-5)          &100 &        &  297   & 555                  &        & 4 (-2)*& 6 (-3)               &                       &\rev{\phantom{0}6.9*} & \rev{2.3}      &\\
5 (-6)          &200 &        &        &1165                  &        &        & 1 (-2)               &                       &                      & \rev{2.5}      &\\
\hline
\end{tabular}
\end{center}
\tablecomments{
Dependence of SSD saturation energy on  $\eta$,  Pm, and resolution. \rev{The rms velocity} $u_{\rm
rms}$ rises from 50 to $130\kms$ between $\dx=1/2\pc$ and \rev{$1/8\pc$} and
varies as much as 25\% for a given $\dx$ so Rm is not fixed for a given Pm.
\rev{Saturated} magnetic energy density $e_{\rm s}$ is averaged over the final 2.5\% of
each run, normalized by the time-averaged turbulent kinetic energy density
$\overline{e_{\rm K}}$.
\rev{Applying
Eq.~\eqref{eq:lB}, the integral scale $L_B$ of the magnetic field is derived from the  energy spectra, aggregated during the
final megayear of each run.}
Asterisks indicate runs where the field appears saturated.
}
\end{table*}
%------------------------------------------------------------------------

\rev{The integral scale of the magnetic field, above which the field becomes weakly
correlated with itself, is
\begin{equation}\label{eq:lB}
L_B=2\uppi\dfrac{\sum_k k^{-1} e_\text{M}(k)}{\sum_k e_\text{M}(k)}
\end{equation}
 in the case of a solenoidal vector\footnote{\rev{\citet[][equation~{[}5{]}, also
\citealt{Hollins22}]{Gent:2013a}  apply the integral scale for an isotropic
\emph{scalar} field, lower by a factor of 4, though that does not alter their
conclusions.}} \citep[Section 12.1,][]{MY07}. To calculate $L_B$, we aggregated the spectra
during the final megayear of models \PCB--\PEE\ (20 snapshots for \PCB\ and 10
snapshots otherwise). We provide the value of $L_B$ in Table~\ref{tab:Rm}.}

\rev{Models in the \PCB\ family have evolved beyond the linear kinematic stage of the SSD, saturating
with $L_B\simeq11\pc$, higher than models~\NCA\, which are still in the kinematic stage, and have
$L_B\simeq6\pc$. Models~\NAA\ and
\NBA\ have $L_B\simeq21$ and $12\pc$, respectively, increasing only for
$\eta\gg\eta_\text{crit}$ and otherwise being quite insensitive to $\eta$. With
model~\PEE\, also still kinematic and having $L_B\simeq2.5\pc$, the integral
scale appears bounded by resolution during the kinematic stage of the SSD.}

\rev{However, models~\PDC\,, all in the nonlinear stage of the SSD, have
integral scales $L_B>3\pc$, which would be expected during the kinematic stage
following the trends evident in models \NAA--\NCA. They may also converge on
$L_B\simeq11\pc$, but confirmation will require the simulations to be extended.
The equivalent integral scale for the velocity is consistently
$L_{U}\simeq121$--$129$~pc with little sensitivity to resolution, hence
$\nu=10^{-3}\kpc\kms$ is \emph{fully} resolved.  This is somewhat smaller than
$L_v = 270\pc$ for the turbulent flow found by \citet{Hollins22}, but that determination was in lower densities entrained by vertical winds and galactic shear.}

The current results at resolutions of $512^3$ and $1024^3$ are strong evidence
that the magnetic energy at saturation of the SSD is asymptotic at Pm or Rm
levels far below those of the real ISM and within reach of numerical
simulations. This is consistent with the results of \citet{SCTMM04} and
\citet{SSFBK15} for isothermal, compressible plasmas with artificial forcing,
who find  at $\Pm\lesssim100$ there is already an asymptotic value of the magnetic saturation energy below 10\% of
equipartition with the kinetic energy.

%--------------------------------------------------------------------------
\section{Summary and conclusions}\label{sec:Sum}
%--------------------------------------------------------------------------

Models with increasing magnetic Prandtl number Pm show that the magnetic energy
density amplified by the SSD in the ISM saturates at about 5\% of the turbulent
kinetic energy density\rev{. This is} significantly below the turbulent magnetic field energy
density estimated from observations.  \rev{Saturation} occurs at relatively low Prandtl
numbers Pm of order 10--100, while estimates of the actual Pm for the ISM are in
the range $10^{10}$--$10^{14}$ \citep{BS05}.

In this study we have not considered the variation of the fluid Reynolds number
$\Rey$, because in previous experiments Rm rather than Pm has been found to be
the determining parameter for the SSD properties
\citep{oishi2011,GMKS21,GMKS22}.  However, these results were at much lower
resolution, so the effect of $\Rey$ should still be tested at high resolution.
\rev{Even then, $\Rey$ also would remain orders of magnitude below the values in
the ISM. Therefore, before such results could be considered robust, we would need to find evidence as $\Rey$ grows of similar asymptotic behavior in key
features of the flow as we find for the magnetic field.}

It is possible that we have a
false convergence in the properties of the SSD \citep{FMA91}, and that at even
higher Rm, new behaviour emerges.  In \citet{GMKS21}, a false convergence of
magnetic decay for $\dx\geq2\pc$ at $\eta=10^{-3}\kpc\kms$ and $\nu=0$ is
contradicted at $\dx\leq1\pc$, for which SSD is present. However, this can be
explained by the numerical diffusion dominating over the physical diffusion,
whereas here we have ensured that $\eta>\eta_{\rm crit}$, so any qualitative
change in the behaviour of the SSD would imply that new or additional physics
are acting at the smaller scales.

In these experiments, we have yet to see convergence of the growth rate in the
SSD, with faster growth at higher Rm and also at higher resolution.
\rev{Affecting its growth, a measure of the integral scale $L_B$ of the
magnetic field during the linear stage of the SSD appears insensitive to Rm,
but not to resolution in the range considered.  In the nonlinear stage, we find
signs of convergence in $L_B$ across resolution and Rm. We may impute this to
be a physical response to the ISM turbulence, but we need to examine how $L_B$
responds to flows with higher Re.}

From our highest resolution runs, we can
verify that in the ISM we would expect a turbulent magnetic
field to reach 5\% of equipartion with the turbulent kinetic energy within less than
1--2 Myr. This is a timescale that is almost negligible for cosmological,
galaxy cluster, or full galaxy simulations, in which it is unfeasible due to the required numerical resolution to produce
by dynamo such strong magnetic fields on
current computational resources.
%It would therefore be worth considering including these turbulent fields as an
%input to such simulations at the outset.  For such models and also for LSD
%models, the SSD would be too slow, and the benefit of our results would be to
%model the characteristics and strength of the turbulent magnetic field either
%by a subgrid-scale model or another method to ensure its effect is present.
It would therefore be worth considering \rev{whether} these turbulent fields
\rev{might be included} as an input to such simulations at the outset.

%lt Turbulent magnetic fields in the ISM are orders of magnitude larger than can be
%produced by the SSD. \citet{GMK24} determine that tangling of the large-scale field while
%the LSD is active is the primary generator of the turbulent field to the
%observed magnitudes, capable of yielding turbulent fields hundreds of times
%stronger than the tangled large-scale field. Perhaps, when Rm is too low, the
%tangling becomes too inefficient to sustain sufficiently strong turbulent
%fields once the LSD approaches saturation and its Lorentz force inhibits the
%turbulent flow. Analysis of these SSD models may provide the insight required
%to sufficiently restore the turbulent constituent.

\rev{SSD production of turbulent magnetic field is absent or under-resolved in
large-scale simulations.  Our results suggest that it is possible to
capture numerically many critical features of this process.  Models of LSD that
partially resolve the SSD \citep{GMK24} help clarify how these interact.  Quantification of this process could be included in the analysis and
interpretation of models in which it is underresolved or absent.  }
%Our magnetic energy spectra suggest that an addition to the simulated magnetic
%field, which exhibits the Kazantsev $k^{3/2}$ scaling at all $k$  could be
%included in an LSD or galactic simulation. \citet{GMK24} observe that the
%spectrum during the kinematic dynamo stage has the same scaling at low $k$, but
%its peak shifting to lower $k$ due to tangling. In this phase, the largest
%scales of the LSD grow most rapidly eventually superseding the Kazantsev
%scaling. So, the energy peak beyond which this scaling disappears does appear
%to move to lower $k$ as the SSD driven field saturates, but to find whether or
%where this becomes asymptotic with increasing Rm would require continuation of
%the models in this work, particularly at the highest resolution, beyond
%saturation of the SSD.  In most practical models of LSD or galaxy evolution, it
%is likely that this peak in any case applies below their grid resolution. From
%this perspective the structure of the turbulent field does not appear to depend
%on whether it is derived by SSD or tangling.  How to add an artificial
%turbulent field and whether to apply it proportional to the energy density of
%the total or the large-scale magnetic field, the turbulent kinetic energy
%density or other parameters requires further analysis.

\facilities{The authors wish to acknowledge CSC – IT Center for Science,
Finland, for computational resources. We acknowledge the EuroHPC Joint
Undertaking for awarding this project access to the EuroHPC supercomputer
LUMI, hosted by CSC (Finland) and the LUMI consortium through a EuroHPC
Extreme Scale Access call.}
\software{\rev{\\{\sc \mbox{Pencil Code}\:}\url{https://pencil-code.nordita.org} \\{\sc
Astaroth\:}\url{https://bitbucket.org/jpekkila/astaroth/src/master}}}
%------------------------------------------------------------------------

\begin{acknowledgments}
\rev{We thank the anonymous referee for constructive criticism that led to
improved quality and presentation of our work.}
We thank C. F. Gammie, C. Federrath, and J. Beattie for useful discussion of
resolution diagnostics and S.  Srivastava \& P Chatterjee for discussion
regarding the energy conservation term in Eq.~\eqref{eq:ent}.  F.A.G.
benefited from in-depth discussions at ``Towards a Comprehensive Model of the
Galactic Magnetic Field'' at Nordita 2023 supported by NordForsk and
acknowledges support of the Swedish Research Council (Vetenskapsrådet) grant
no. 2022–03767.  M.J.K.-L., M.R. and T.P. acknowledge support from the ERC
under the EU's Horizon 2020 research and innovation programme (Project UniSDyn,
grant 818665). M-M.M.L. was partly supported by US NSF grant AST23-07950, and
thanks the Inst.\ f\"ur Theoretische Astrophysik der Uni.\ Heidelberg for
hospitality.  His research was also supported in part by grant NSF PHY-2309135
to the Kavli Institute for Theoretical Physics (KITP).
\end{acknowledgments}
%------------------------------------------------------------------------

%----------------------------------------------------------------------------
\bibliographystyle{aasjournal}
\bibliography{refs}{}

@article{Pencil-JOSS,
       author = {{Pencil Code Collaboration} and {Brandenburg}, Axel and {Johansen}, Anders and {Bourdin}, Philippe and {Dobler}, Wolfgang and {Lyra}, Wladimir and {Rheinhardt}, Matthias and {Bingert}, Sven and {Haugen}, Nils and {Mee}, Antony and {Gent}, Frederick and {Babkovskaia}, Natalia and {Yang}, Chao-Chin and {Heinemann}, Tobias and {Dintrans}, Boris and {Mitra}, Dhrubaditya and {Candelaresi}, Simon and {Warnecke}, J{\"o}rn and {K{\"a}pyl{\"a}}, Petri and {Schreiber}, Andreas and {Chatterjee}, Piyali and {K{\"a}pyl{\"a}}, Maarit and {Li}, Xiang-Yu and {Kr{\"u}ger}, Jonas and {Aarnes}, J{\o}rgen and {Sarson}, Graeme and {Oishi}, Jeffrey and {Schober}, Jennifer and {Plasson}, Rapha{\"e}l and {Sandin}, Christer and {Karchniwy}, Ewa and {Rodrigues}, Luiz and {Hubbard}, Alexander and {Guerrero}, Gustavo and {Snodin}, Andrew and {Losada}, Illa and {Pekkil{\"a}}, Johannes and {Qian}, Chengeng},
        title = "{The Pencil Code, a modular MPI code for partial differential equations and particles: multipurpose and multiuser-maintained}",
      journal = {The Journal of Open Source Software},
         year = 2021,
        month = feb,
       volume = {6},
       number = {58},
          eid = {2807},
        pages = {2807},
          doi = {10.21105/joss.02807},
archivePrefix = {arXiv},
       eprint = {2009.08231},
 primaryClass = {astro-ph.IM},
       adsurl = {https://ui.adsabs.harvard.edu/abs/2021JOSS....6.2807P}
}

@ARTICLE{Beck15,
       author = {{Beck}, Rainer},
        title = "{Magnetic fields in spiral galaxies}",
      journal = {\aapr},
         year = 2015,
        month = dec,
       volume = {24},
          eid = {4},
        pages = {4},
          doi = {10.1007/s00159-015-0084-4},
archivePrefix = {arXiv},
       eprint = {1509.04522},
 primaryClass = {astro-ph.GA},
       adsurl = {https://ui.adsabs.harvard.edu/abs/2015A&ARv..24....4B}
}

@phdthesis{Gent:2012,
  author   ={Frederick A. Gent},
  title    ={Supernova Driven Turbulence in the Interstellar Medium},
  year     ={2012},
  month  = {November},
  school   ={Newcastle University School of Mathematics and Statistics},
  howpublished  =  "\url{http://hdl.handle.net/10443/1755}",
  eprint         =   {http://hdl.handle.net/10443/1755}
}

@ARTICLE{Gent:2013a,
   author = {{Gent}, F.~A. and {Shukurov}, A. and {Sarson}, G.~R. and {Fletcher}, A. and
	{Mantere}, M.~J.},
    title = "{The supernova-regulated ISM - II. The mean magnetic field}",
  journal = {\mnras},
archivePrefix = "arXiv",
   eprint = {1206.6784},
 primaryClass = "astro-ph.GA",
     year = 2013,
    month = mar,
   volume = 430,
    pages = {L40-L44},
      doi = {10.1093/mnrasl/sls042},
   adsurl = {http://adsabs.harvard.edu/abs/2013MNRAS.430L..40G}
}

@ARTICLE{Gent:2013b,
   author = {{Gent}, F.~A. and {Shukurov}, A. and {Fletcher}, A. and {Sarson}, G.~R. and
	{Mantere}, M.~J.},
    title = "{The supernova-regulated ISM - I. The multiphase structure}",
  journal = {\mnras},
archivePrefix = "arXiv",
   eprint = {1204.3567},
 primaryClass = "astro-ph.GA",
     year = 2013,
    month = jun,
   volume = 432,
    pages = {1396-1423},
      doi = {10.1093/mnras/stt560},
   adsurl = {http://adsabs.harvard.edu/abs/2013MNRAS.432.1396G}
}

@ARTICLE{Gressel:2008,
   author = {{Gressel}, O. and {Ziegler}, U. and {Elstner}, D. and {R{\"u}diger}, G.
	},
    title = "{Dynamo coefficients from local simulations of the turbulent ISM}",
  journal = {Astronomische Nachrichten},
archivePrefix = "arXiv",
   eprint = {0801.4004},
     year = 2008,
    month = jul,
   volume = 329,
    pages = {619},
      doi = {10.1002/asna.200811005},
   adsurl = {http://adsabs.harvard.edu/abs/2008AN....329..619G}
}

@ARTICLE{K68,
       author = {{Kazantsev}, A.~P.},
        title = "{Enhancement of a Magnetic Field by a Conducting Fluid}",
      journal = {Soviet Journal of Experimental and Theoretical Physics},
         year = 1968,
        month = may,
       volume = {26},
        pages = {1031},
       adsurl = {https://ui.adsabs.harvard.edu/abs/1968JETP...26.1031K}
}

@ARTICLE{Wolfire:1995,
   author = {{Wolfire}, M.~G. and {Hollenbach}, D. and {McKee}, C.~F. and
	{Tielens}, A.~G.~G.~M. and {Bakes}, E.~L.~O.},
    title = "{The neutral atomic phases of the interstellar medium}",
  journal = {\apj},
     year = 1995,
    month = apr,
   volume = 443,
    pages = {152-168},
      doi = {10.1086/175510},
   adsurl = {http://adsabs.harvard.edu/abs/1995ApJ...443..152W}
}

@ARTICLE{GMKSH20,
       author = {{Gent}, F.~A. and {Mac Low}, M.-M. and {K{\"a}pyl{\"a}}, M.~J. and
         {Sarson}, G.~R. and {Hollins}, J.~F.},
        title = "{Modelling supernova-driven turbulence}",
      journal = {Geophysical and Astrophysical Fluid Dynamics},
         year = 2020,
        month = mar,
       volume = {114},
       number = {1-2},
        pages = {77-105},
          doi = {10.1080/03091929.2019.1634705},
archivePrefix = {arXiv},
       eprint = {1806.01570},
 primaryClass = {astro-ph.GA},
       adsurl = {https://ui.adsabs.harvard.edu/abs/2020GApFD.114...77G}
}

@ARTICLE{FMA91,
       author = {{Fryxell}, Bruce and {Mueller}, Ewald and {Arnett}, David},
        title = "{Instabilities and Clumping in SN 1987A. I. Early Evolution in Two Dimensions}",
      journal = {\apj},
         year = 1991,
        month = feb,
       volume = {367},
        pages = {619},
          doi = {10.1086/169657},
       adsurl = {https://ui.adsabs.harvard.edu/abs/1991ApJ...367..619F}
}

@ARTICLE{HB04,
       author = {{Haugen}, Nils Erland L. and {Brandenburg}, Axel},
        title = "{Suppression of small scale dynamo action by an imposed magnetic field}",
      journal = {\pre},
         year = 2004,
        month = sep,
       volume = {70},
       number = {3},
          eid = {036408},
        pages = {036408},
          doi = {10.1103/PhysRevE.70.036408},
archivePrefix = {arXiv},
       eprint = {astro-ph/0402281},
 primaryClass = {astro-ph},
       adsurl = {https://ui.adsabs.harvard.edu/abs/2004PhRvE..70c6408H}
}

@ARTICLE{HSSFG17,
   author = {{Hollins}, J.~F. and {Sarson}, G.~R. and {Shukurov}, A. and
	{Fletcher}, A. and {Gent}, F.~A.},
    title = "{Supernova-regulated ISM. V. Space and Time Correlations}",
  journal = {\apj},
archivePrefix = "arXiv",
   eprint = {1703.05187},
     year = 2017,
    month = nov,
   volume = 850,
      eid = {4},
    pages = {4},
      doi = {10.3847/1538-4357/aa93e7},
   adsurl = {http://adsabs.harvard.edu/abs/2017ApJ...850....4H}
}

@ARTICLE{RT16,
       author = {{Rieder}, Michael and {Teyssier}, Romain},
        title = "{A small-scale dynamo in feedback-dominated galaxies as the origin of cosmic magnetic fields - I. The kinematic phase}",
      journal = {\mnras},
         year = 2016,
        month = apr,
       volume = {457},
       number = {2},
        pages = {1722-1738},
          doi = {10.1093/mnras/stv2985},
archivePrefix = {arXiv},
       eprint = {1506.00849},
 primaryClass = {astro-ph.GA},
       adsurl = {https://ui.adsabs.harvard.edu/abs/2016MNRAS.457.1722R}
}

@ARTICLE{RT17,
       author = {{Rieder}, Michael and {Teyssier}, Romain},
        title = "{A small-scale dynamo in feedback-dominated galaxies - II. The saturation phase and the final magnetic configuration}",
      journal = {\mnras},
         year = 2017,
        month = nov,
       volume = {471},
       number = {3},
        pages = {2674-2686},
          doi = {10.1093/mnras/stx1670},
archivePrefix = {arXiv},
       eprint = {1704.05845},
 primaryClass = {astro-ph.GA},
       adsurl = {https://ui.adsabs.harvard.edu/abs/2017MNRAS.471.2674R}
}

@ARTICLE{RT17a,
       author = {{Rieder}, Michael and {Teyssier}, Romain},
        title = "{A small-scale dynamo in feedback-dominated galaxies - III. Cosmological simulations}",
      journal = {\mnras},
         year = 2017,
        month = dec,
       volume = {472},
       number = {4},
        pages = {4368-4373},
          doi = {10.1093/mnras/stx2276},
archivePrefix = {arXiv},
       eprint = {1708.01486},
 primaryClass = {astro-ph.GA},
       adsurl = {https://ui.adsabs.harvard.edu/abs/2017MNRAS.472.4368R}
}

@phdthesis{Gressel08b,
  author   =  "Oliver Gressel",
  title    =  "Supernova-driven Turbulence and Magnetic Field Amplification in Disk Galaxies",
  year     =  "2008",
  school   =  "Astrophysikalisches Institut Potsdam"
}

@article{oishi2011,
	author = {{Oishi}, J.~S. and {Mac Low}, M.-M.},
	journal = {\apj},
	month = oct,
	pages = {18},
	title = {{Magnetorotational Turbulence Transports Angular Momentum in Stratified Disks with Low Magnetic Prandtl Number but Magnetic Reynolds Number above a Critical Value}},
	volume = 740,
	year = 2011}

@ARTICLE{GMKS21,
       author = {{Gent}, Frederick A. and {Mac Low}, Mordecai-Mark and {K{\"a}pyl{\"a}}, Maarit J. and {Singh}, Nishant K.},
       title = "{Small-scale Dynamo in Supernova-driven Interstellar Turbulence}",
     journal = {\apjl},
   year = 2021,
   month = apr,
  volume = {910},
 number = {2},
   eid = {L15},
   pages = {L15},
     doi = {10.3847/2041-8213/abed59},
     archivePrefix = {arXiv},
    eprint = {2010.01833},
     primaryClass = {astro-ph.GA},
    adsurl = {https://ui.adsabs.harvard.edu/abs/2021ApJ...910L..15G}
}

@ARTICLE{KO15,
       author = {{Kim}, Chang-Goo and {Ostriker}, Eve C.},
        title = "{Momentum Injection by Supernovae in the Interstellar Medium}",
      journal = {\apj},
         year = 2015,
        month = apr,
       volume = {802},
       number = {2},
          eid = {99},
        pages = {99},
          doi = {10.1088/0004-637X/802/2/99},
archivePrefix = {arXiv},
       eprint = {1410.1537},
 primaryClass = {astro-ph.GA},
       adsurl = {https://ui.adsabs.harvard.edu/abs/2015ApJ...802...99K}
}

@ARTICLE{SSFBK15,
       author = {{Schober}, J. and {Schleicher}, D.~R.~G. and {Federrath}, C. and {Bovino}, S. and {Klessen}, R.~S.},
        title = "{Saturation of the turbulent dynamo}",
      journal = {\pre},
         year = 2015,
        month = aug,
       volume = {92},
       number = {2},
          eid = {023010},
        pages = {023010},
          doi = {10.1103/PhysRevE.92.023010},
archivePrefix = {arXiv},
       eprint = {1506.02182},
 primaryClass = {physics.plasm-ph},
       adsurl = {https://ui.adsabs.harvard.edu/abs/2015PhRvE..92b3010S}
}

@ARTICLE{SVG02,
       author = {{S{\'a}nchez-Salcedo}, F.~J. and {V{\'a}zquez-Semadeni}, E. and {Gazol}, A.},
        title = "{The Nonlinear Development of the Thermal Instability in the Atomic Interstellar Medium and Its Interaction with Random Fluctuations}",
      journal = {\apj},
         year = 2002,
        month = oct,
       volume = {577},
       number = {2},
        pages = {768-788},
          doi = {10.1086/342223},
archivePrefix = {arXiv},
       eprint = {astro-ph/0203067},
 primaryClass = {astro-ph},
       adsurl = {https://ui.adsabs.harvard.edu/abs/2002ApJ...577..768S}
}

@ARTICLE{GMKS22,
       author = {{Gent}, Frederick A. and {Mac Low}, Mordecai-Mark and {Korpi-Lagg}, Maarit J. and {Singh}, Nishant K.},
        title = "{The Small-scale Dynamo in a Multiphase Supernova-driven Medium}",
      journal = {\apj},
         year = 2023,
        month = feb,
       volume = {943},
       number = {2},
          eid = {176},
        pages = {176},
          doi = {10.3847/1538-4357/acac20},
archivePrefix = {arXiv},
       eprint = {2210.04460},
 primaryClass = {astro-ph.GA},
       adsurl = {https://ui.adsabs.harvard.edu/abs/2023ApJ...943..176G}
}

@ARTICLE{BT91,
       author = {{van den Bergh}, Sidney and {Tammann}, Gustav A.},
        title = "{Galactic and extragalactic supernova rates.}",
      journal = {\araa},
         year = 1991,
        month = jan,
       volume = {29},
        pages = {363-407},
          doi = {10.1146/annurev.aa.29.090191.002051},
       adsurl = {https://ui.adsabs.harvard.edu/abs/1991ARA&A..29..363V}
}

@ARTICLE{Mannucci05,
       author = {{Mannucci}, F. and {Della Valle}, M. and {Panagia}, N. and {Cappellaro}, E. and {Cresci}, G. and {Maiolino}, R. and {Petrosian}, A. and {Turatto}, M.},
        title = "{The supernova rate per unit mass}",
      journal = {\aap},
         year = 2005,
        month = apr,
       volume = {433},
       number = {3},
        pages = {807-814},
          doi = {10.1051/0004-6361:20041411},
archivePrefix = {arXiv},
       eprint = {astro-ph/0411450},
 primaryClass = {astro-ph},
       adsurl = {https://ui.adsabs.harvard.edu/abs/2005A&A...433..807M}
}

@ARTICLE{BS05,
       author = {{Brandenburg}, Axel and {Subramanian}, Kandaswamy},
        title = "{Astrophysical magnetic fields and nonlinear dynamo theory}",
      journal = {\physrep},
         year = 2005,
        month = oct,
       volume = {417},
       number = {1-4},
        pages = {1-209},
          doi = {10.1016/j.physrep.2005.06.005},
archivePrefix = {arXiv},
       eprint = {astro-ph/0405052},
 primaryClass = {astro-ph},
       adsurl = {https://ui.adsabs.harvard.edu/abs/2005PhR...417....1B}
}

@ARTICLE{GMK24,
       author = {{Gent}, Frederick A. and {Mac Low}, Mordecai-Mark and {Korpi-Lagg}, Maarit J.},
        title = "{Transition from Small-scale to Large-scale Dynamo in a Supernova-driven, Multiphase Medium}",
      journal = {\apj},
         year = 2024,
        month = jan,
       volume = {961},
       number = {1},
          eid = {7},
        pages = {7},
          doi = {10.3847/1538-4357/ad0da0},
archivePrefix = {arXiv},
       eprint = {2306.07051},
 primaryClass = {astro-ph.GA},
       adsurl = {https://ui.adsabs.harvard.edu/abs/2024ApJ...961....7G}
}

@ARTICLE{BB25,
       author = {{Beck}, R. and {Berkhuijsen}, E.~M.},
        title = "{Magnetic fields and cosmic rays in M 31: II. Strength and distribution of the magnetic field components}",
      journal = {\aap},
         year = 2025,
        month = aug,
       volume = {700},
          eid = {A198},
        pages = {A198},
          doi = {10.1051/0004-6361/202555048},
archivePrefix = {arXiv},
       eprint = {2507.07719},
 primaryClass = {astro-ph.GA},
       adsurl = {https://ui.adsabs.harvard.edu/abs/2025A&A...700A.198B}
}

@ARTICLE{Tevlin25,
       author = {{Tevlin}, L. and {Berlok}, T. and {Pfrommer}, C. and {Talbot}, R.~Y. and {Whittingham}, J. and {Puchwein}, E. and {Pakmor}, R. and {Weinberger}, R. and {Springel}, V.},
        title = "{Magnetic dynamos in galaxy clusters: The crucial role of galaxy formation physics at high redshifts}",
      journal = {\aap},
         year = 2025,
        month = sep,
       volume = {701},
          eid = {A114},
        pages = {A114},
          doi = {10.1051/0004-6361/202452823},
archivePrefix = {arXiv},
       eprint = {2411.00103},
 primaryClass = {astro-ph.GA},
       adsurl = {https://ui.adsabs.harvard.edu/abs/2025A&A...701A.114T}
}

@ARTICLE{Kriel22,
       author = {{Kriel}, Neco and {Beattie}, James R. and {Seta}, Amit and {Federrath}, Christoph},
        title = "{Fundamental scales in the kinematic phase of the turbulent dynamo}",
      journal = {\mnras},
         year = 2022,
        month = jun,
       volume = {513},
       number = {2},
        pages = {2457-2470},
          doi = {10.1093/mnras/stac969},
archivePrefix = {arXiv},
       eprint = {2204.00828},
 primaryClass = {astro-ph.SR},
       adsurl = {https://ui.adsabs.harvard.edu/abs/2022MNRAS.513.2457K}
}

@ARTICLE{Kriel25,
       author = {{Kriel}, Neco and {Beattie}, James R. and {Federrath}, Christoph and {Krumholz}, Mark R. and {Hew}, Justin Kin Jun},
        title = "{Fundamental MHD scales - II. The kinematic phase of the supersonic small-scale dynamo}",
      journal = {\mnras},
         year = 2025,
        month = mar,
       volume = {537},
       number = {3},
        pages = {2602-2629},
          doi = {10.1093/mnras/staf188},
archivePrefix = {arXiv},
       eprint = {2310.17036},
 primaryClass = {astro-ph.GA},
       adsurl = {https://ui.adsabs.harvard.edu/abs/2025MNRAS.537.2602K}
}

@article{Pekkila19,
  title={{Astaroth: A library for stencil computations on graphics processing units}},
  author={Pekkil{\"a}, Johannes},
  year={2019},
  url={https://aaltodoc.aalto.fi/server/api/core/bitstreams/c73ad7b3-47a2-4c23-b802-7721366fb961/content}
}

@article{Pekkila22,
  title={{Scalable communication for high-order stencil computations using CUDA-aware MPI}},
  author={Pekkil{\"a}, Johannes and V{\"a}is{\"a}l{\"a}, Miikka S and K{\"a}pyl{\"a}, Maarit J and Rheinhardt, Matthias and Lappi, Oskar},
  journal={Parallel Computing},
  volume={111},
  pages={102904},
  year={2022},
  publisher={Elsevier}
}

@ARTICLE{Sankalp25,
       author = {{Srivastava}, Sankalp and {Chatterjee}, Piyali and {Dey}, Sahel and {Erd{\'e}lyi}, Robertus},
        title = "{The Relation between Solar Spicules and Magnetohydrodynamic Shocks}",
      journal = {\apj},
         year = 2025,
        month = aug,
       volume = {989},
       number = {1},
          eid = {39},
        pages = {39},
          doi = {10.3847/1538-4357/ade9b4},
archivePrefix = {arXiv},
       eprint = {2506.21517},
 primaryClass = {astro-ph.SR},
       adsurl = {https://ui.adsabs.harvard.edu/abs/2025ApJ...989...39S}
}

@ARTICLE{SCTMM04,
       author = {{Schekochihin}, Alexander A. and {Cowley}, Steven C. and {Taylor}, Samuel F. and {Maron}, Jason L. and {McWilliams}, James C.},
        title = "{Simulations of the Small-Scale Turbulent Dynamo}",
      journal = {\apj},
         year = 2004,
        month = sep,
       volume = {612},
       number = {1},
        pages = {276-307},
          doi = {10.1086/422547},
archivePrefix = {arXiv},
       eprint = {astro-ph/0312046},
 primaryClass = {astro-ph},
       adsurl = {https://ui.adsabs.harvard.edu/abs/2004ApJ...612..276S}
}

@article{Kennedy00,
  title={Low-storage, Explicit Runge-Kutta Schemes for the Compressible Navier-Stokes Equations},
  author={Christopher A. Kennedy and Mark H. Carpenter and R. Michael Lewis},
  journal={Applied Numerical Mathematics},
  year={2000},
  volume={35},
  pages={177-219},
  url={https://api.semanticscholar.org/CorpusID:15438200}
}

@article{V21,
doi = {10.3847/1538-4357/abceca},
url = {https://doi.org/10.3847/1538-4357/abceca},
year = {2021},
month = {feb},
publisher = {The American Astronomical Society},
volume = {907},
number = {2},
pages = {83},
author = {{Väisälä}, Miikka S. and Pekkilä, Johannes and Käpylä, Maarit J. and Rheinhardt, Matthias and Shang, Hsien and Krasnopolsky, Ruben},
title = {Interaction of Large- and Small-scale Dynamos in Isotropic Turbulent Flows from GPU-accelerated Simulations},
journal = {The Astrophysical Journal},
}

@inbook{MY07,
author = {Monin, A. and Yaglom, A.},
year = {2007},
month = {01},
publisher = {M.I.T. Press},
pages = {},
title = {Statistical Fluid Mechanics, Volume 2: Mechanics of Turbulence},
isbn = {9780486458915}
}

@ARTICLE{Hollins22,
       author = {{Hollins}, James F. and {Sarson}, Graeme R. and {Evirgen}, Cetin Can and {Shukurov}, Anvar and {Fletcher}, Andrew and {Gent}, Frederick A.},
        title = "{Mean fields and fluctuations in compressible magnetohydrodynamic flows}",
      journal = {Geophysical and Astrophysical Fluid Dynamics},
         year = 2022,
        month = jul,
       volume = {116},
       number = {4},
        pages = {261-289},
          doi = {10.1080/03091929.2022.2032022},
archivePrefix = {arXiv},
       eprint = {1809.01098},
 primaryClass = {astro-ph.GA},
       adsurl = {https://ui.adsabs.harvard.edu/abs/2022GApFD.116..261H}
}
%----------------------------------------------------------------------------

%----------------------------------------------------------------------------
\end{document}